\documentclass[10pt,onecolumn]{article}
\usepackage{geometry}                
\usepackage{graphicx}
\usepackage{amssymb}
\usepackage{epstopdf}
\usepackage{array}
\usepackage{amsthm}
\usepackage{multirow}
\usepackage{amsmath}

\usepackage{wrapfig}
\usepackage{bm}

\theoremstyle{plain}
\newcommand{\argmax}[1]{\underset{#1}{\operatorname{arg}\!\operatorname{max}}\;}

\theoremstyle{definition}

\usepackage{authblk}
\title{Multi-agent Economics and the Emergence of Critical Markets}  
\author[1]{Michael S. Harr\'e}
\affil[1]{Complex Systems Research Group,
Faculty of Engineering and IT, The University of Sydney, Sydney, Australia.}
\date{}                     
\begin{document}
\maketitle

\begin{abstract}
The dual crises of the sub-prime mortgage crisis and the global financial crisis has prompted a call for explanations of non-equilibrium market dynamics. Recently a promising approach has been the use of agent based models (ABMs) to simulate aggregate market dynamics. A key aspect of these models is the endogenous emergence of critical transitions between equilibria, i.e. market collapses, caused by multiple equilibria and changing market parameters. Several research themes have developed microeconomic based models that include multiple equilibria: social decision theory (Brock and Durlauf), quantal response models (McKelvey and Palfrey), and strategic complementarities (Goldstein). A gap that needs to be filled in the literature is a unified analysis of the relationship between these models and how aggregate criticality emerges from the individual agent level. This article reviews the agent-based foundations of markets starting with the individual agent perspective of McFadden and the aggregate perspective of catastrophe theory emphasising connections between the different approaches. It is shown that changes in the uncertainty agents have in the value of their interactions with one another, even if these changes are one-sided, plays a central role in systemic market risks such as market instability and the twin crises effect. These interactions can endogenously cause crises that are an emergent phenomena of markets.
\end{abstract}

\section{Introduction}

Multi-agent models have grown in popularity~\cite{aoki1998simple,arthur2006out,tesfatsion2006agent,axtell2007economic,farmer2009economy,gatti2010complex} as a way in which to simulate complex market dynamics that might have no closed form solutions. This has prompted a call~\cite{farmer2009economy,buchanan2009economics,johnson2011financial} for their general use in understanding markets and economies in crisis. Recent work using agent-based models of large markets has looked at the macro-market dynamics for Washington DC~\cite{geanakoplos2012getting} and the European economy~\cite{gencceragent,deissenberg2008eurace}. A multi-agent perspective has been suggested as a predictor of market crises~\cite{caiani2016agent} by using the stock and flow of funds at the balance sheet (agent) level as in Lavoie and Godley~\cite{lavoie2001kaleckian,godley2006monetary}, see for example Bezemer~\cite{bezemer2010understanding,bezemer2012economy} and the agent based modelling approach in~\cite{gualdi2015tipping}. Connections between the multi-agent approach and econometrics have also been studied~\cite{brock2007identification,durlauf2010social} in which aggregate market signals are compared with agent-based models of microeconomic interactions. Connecting these different approaches with Gallegati {\it et al}'s~\cite{gallegati2011period} work on the `period of financial distress' that may be an indicator of a market at risk of collapse would be a significant step in understanding market crises. \\

The agent-based approaches that have either social utilities or game theory interactions between the agents plays a central role in the emergence of systemic risks in markets. Gintis~\cite{gintis2009bounds} has emphasised the unifying role that game theory should play in the unification of the behavioural sciences and Tesfatsion~\cite{tesfatsion2006agent,tesfatsion2001introduction} has emphasised a constructivist approach to non-linearities, for example the early work of Arthur~\cite{arthur1994inductive,arthur2006out,arthur2018asset} in asset markets and Axelrod~\cite{axelrod1981evolution,axelrod1997complexity} in game theory. Specifically these models do not presuppose the existence of an equilibrium that plays a central role in economics~\cite{farmer2009virtues}. So while the emphasis in this article will be on bifurcations other non-linear models of generalised heterogeneous agents~\cite{brock1998heterogeneous,hommes2006heterogeneous} and housing markets specifically~\cite{dieci2012simple} have demonstrated how chaos can emerge from otherwise simple models.  \\

Agent-based models are not all the same in their approach to equilibria and the merits of equilibrium based approaches are covered in Farmer and Geanakoplos~\cite{farmer2009virtues}. Equilibrium models can be usefully divided into three categories: the locally unique equilibrium models developed by Debreu, partial equilibrium models in which equilibrium and non-equilibrium states co-exist in the same model, and non-equilibrium models where equilibria are not assumed and may not exist. Locally unique models distinguish between regular and critical economies~\cite{debreu1970economies,debreu1974four} and assume that regular economies (locally stable and unique) are sufficiently common to ignore critical states. Partial equilibrium models such as catastrophe theory~\cite{casti1976catastrophe,balasko1978economic,barunik2015realizing} allow for equilibrium states but the focus is on non-equilibrium transitions at critical points. This is the approach of the current article. Non-equilibrium models do not assume an equilibrium exists~\cite{axelrod1997complexity,lavoie2001kaleckian,tesfatsion2006agent,geanakoplos2012getting}, the agents often have bounded cognitive abilities (memory, predictive power, multiple predictive models) and are heterogeneous both in their description and the role they can play in a market. The partial equilibrium approach allows us to study what drives a minimal model of non-linear dynamics and draw insights for developing more complex models. A  new approach that bridges the gap between partial equilibrium models and models that exhibit realistic aggregate market dynamics has been developed by Gallegati {\it et al}~\cite{gallegati2011period} in which herding, financial distress, and the distribution of wealth all play an important role. This is covered in Section~\ref{discussion}.


\section{Background}

The focus of this article is understanding the micro and macro basis of `critical markets', those markets that are near (or at) a tipping point at which a market is at risk of a non-equilibrium transition to a new equilibrium state. A promising approach is that of singularity theory~\cite[Introduction]{arnol1992catastrophe} in which qualitative changes in a system occur as the number of equilibrium points change. The analysis of statistical distributions with multiple equilibria can be found in the earlier work of Cobb~\cite{cobb1978stochastic} and later Wagenmakers {\it et al}~\cite{wagenmakers2005transformation} and empirical methods have been implemented in software~\cite{grasman2009fitting} while applications of this method to aggregate economic time series have recently been implemented for housing markets~\cite{bolt2014identifying,diks2016can}. This review connects an important subset of the issues covered in earlier work as they relate to critical markets and their subsequent collapse. \\

The risks of critical markets can be illustrated with a non-linear regression. The model is defined by an $n^{\mathrm{th}}$ order polynomial of a dynamic variable $x_t$ and parametrised by a vector of coefficients\footnote{These dynamics are put in the context of a market in the Discussion, Section~\ref{discussion}}: 
\begin{eqnarray}
m & = & [\alpha_0, \alpha_1, \ldots , \alpha_n] \in \mathbb{R}^{n+1}.
\end{eqnarray} 
The model predicts the outcome of $x_t$ based on the previous state $x_{t-1}$: 
\begin{eqnarray}
x_t & = & \alpha_0 + \alpha_1 x_{t-1} + ... + \alpha_n x_{t-1}^n\pm \epsilon \,\, = \,\, \mathcal{M}(x_{t-1}\,|\,m). \label{first_eg}
\end{eqnarray} 
The function $x_t = \mathcal{M}(x_{t-1}\,|\,m)$ has two sources of uncertainty: errors in estimates of $m$ and errors in the prediction $x_t$. The first type of error changes the market being described to that of a nearby market, if:
\begin{eqnarray*} 
m + \delta m_i & = & [\alpha_0, \ldots, \alpha_i + \delta \alpha_i, \ldots, \alpha_n]
\end{eqnarray*} 
for small constant $\delta$ then $m + \delta m_i$ describes a market nearby to $m$ and $m \pm \delta m$ will simply mean the aggregate uncertainty in market parameters. The second type of error are fluctuations in the prediction of $x_t$: if $x^*$ is a time independent stationary solution of a market $m$ then the equilibrium $x^* =$ E$(\mathcal{M}(x^*\,|\,m))$ has statistical fluctuations $ x^* \pm \delta x^*$ caused by $\pm\epsilon$. \\
\begin{figure}[!ht]
\center 
\includegraphics[width=.9\columnwidth]{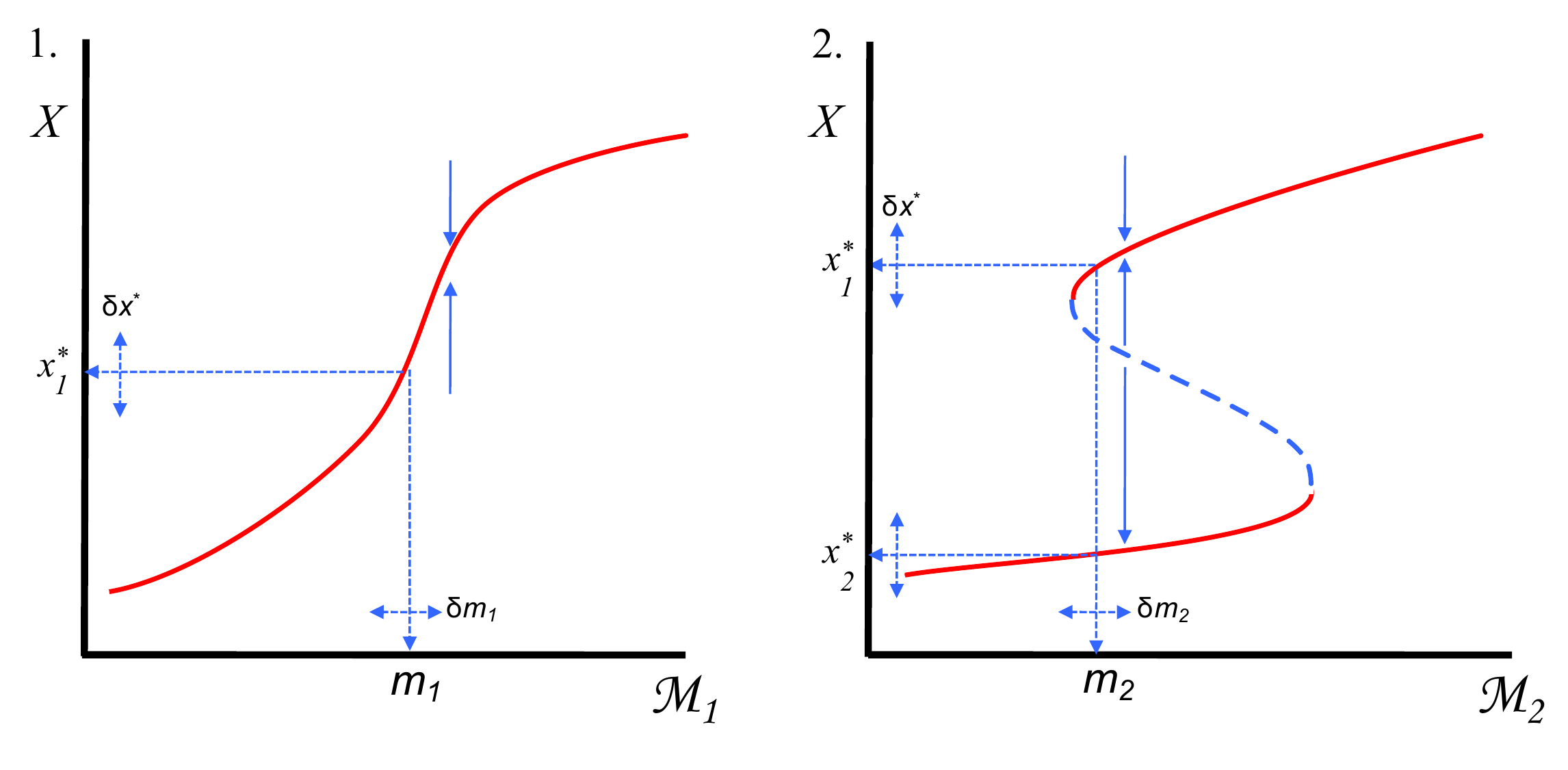}
\label{Structural_Risk}
\caption{Two models $\mathcal{M}_i(x\,|\,m_i)$ and their equilibrium surfaces E$(\mathcal{M}_i(x^*_j\,|\,m_i))=x_j^*$: Plot 1 has a single unique equilibrium $x^*_1$ and there are no systemic risks for any description $m_1$. Plot 2 is the more common situation in which a model has multiple, distinct equilibria $x^*_1$ and $x^*_2$ for some descriptions $m_2$. Critical markets are the $m_2$ at which the number of equilibria change.}
\end{figure}

The systemic risks and market criticality are illustrated in Figure~\ref{Structural_Risk} where two different models $\mathcal{M}_1$ and $\mathcal{M}_2$ are described by the parameter sets $m_1$ and $m_2$ (the $m_i$ in the figures are single valued representations of multivalued vectors) and equilibria are indexed by $j$: $x_j^* =$ E($\mathcal{M}_i(x_j^*\,|\,m_i))$. Plot 1 shows a systemically risk-free market $\mathcal{M}_1$ that has a globally unique, stable equilibrium for every $m_1$ where small variations in the equilibrium state $x_1^* \pm \delta x^*$ or the market's description $m_1 \pm \delta m$ never results in a qualitative change in the location of the equilibrium. \\

Plot 2 is the more common case in which there are multiple locally unique equilibria for some values of $m_2$. By `locally unique' it is meant~\cite{debreu1974four} that for sufficiently small variations $x_1^* \pm \delta x^*$ or $m_2 \pm \delta m$ the market returns to either the same or a nearby equilibrium. However there are two types of systemic risk present in Plot 2 that are not present in Plot 1. Near a bifurcation point a small $\delta x_1^*$ results in $x_1$ passing over a threshold value (the unstable equilibrium, dashed blue line) and goes through a non-equilibrium transition to the substantively different equilibrium at $x^*_2$. Alternatively, for a small change in $m_2$ the market will become critical before collapsing to a substantively different equilibrium near $x^*_2$. \\

The study of such critical transitions was stimulated by the work of Thom who developed a theory of singularities called catastrophe theory~\cite{thom1969topological} describing the abrupt qualitative changes in dynamical systems. Specifically Thom was interested in the generic stability of the singularities of potential functions $\mathcal{M}(x\,|\,m): \mathbb{R}^{i \times j} \rightarrow \mathbb{R}^{i}$ that describe the state of a system where $i$ is the dimension of the state variables and $j$ is the co-dimension of the control parameters. A key result is that catastrophes can be classified into seven `elementary catastrophes' and these are a stable, generic property of systems with a maximum dimension of two and a maximum co-dimension of four. More details follow in Section~\ref{section_background}. Early on the connection between economics and catastrophe theory was recognised by economists such as Balasko~\cite{balasko1978economic} and Dierker~\cite{dierker1982regular}. \\

Catastrophe theory has found many applications in economics having been used in financial markets~\cite{zeeman1974unstable}, urban housing dynamics~\cite{casti1976catastrophe,wilson2011catastrophe},\ and housing markets~\cite{diks2016can,bolt2014identifying}. Early on though there were significant criticisms of the excessive and at times inappropriate application of Thom's theory to systems, including economics, in which it was argued the approach was not warranted. Rosser~\cite{rosser2007rise} has reviewed the criticisms relating specifically to economics and concludes that the application of catastrophe theory, and the broader work on singularity theory, can be a productive approach to economic analysis. The applicability of singularity theory in applied economics has recently progressed with statistical techniques for detecting catastrophes in time series~\cite{wagenmakers2005transformation,grasman2009fitting} and their application to market dynamics~\cite{barunik2009can,barunik2015realizing,diks2016can}. \\

This article reviews and develops a theoretical foundation that connects the macro perspective of bifurcations and catastrophe theory to the micro perspective of agent decisions in the tradition of Lucas' critique of macroeconomics~\cite{brunner1983econometric,gatti2010complex,battiston2016complexity}. Section~\ref{section_background} reviews the background on catastrophe theory and game theory, introducing a qualitative connection between the structural changes of catastrophe theory and bifurcations of Nash equilibria. Section~\ref{section_decision_theory} introduces three approaches to microeconomic markets: McFadden's approach~\cite{mcfadden1973conditional}, the social decision theory of Brock and Durlauf~\cite{brock2001discrete}, and the Quantal Response Equilibrium of McKelvey and Palfrey~\cite{mckelvey1995quantal}. McFadden's approach has no interactions between agents and market nonlinearities stem from nonlinear utility functions. Brock and Durlauf's approach extends McFadden to agents in an endogenous, homogeneous `social field' in which nonlinearities stem from an additive social utility. McKelvey and Palfrey is shown to be similar to Brock and Durlauf where the social field is replaced with explicit pairwise interactions between agents who can have different utility functions. All three models include a stochastic heterogeneity in the utility function for each agent. Section~\ref{section_comp_analysis} is a comparative analysis between models, the emergence of market criticality, and the twin crises effect. The final section, Section~\ref{discussion}, discusses the time dependent evolution of Gallegati {\it et al}~\cite{gallegati2011period} and its relationship to market crises and the potential games of Sandholm~\cite{sandholm2009large}.

\begin{figure}[!ht]
\center
\includegraphics[width=.8\columnwidth]{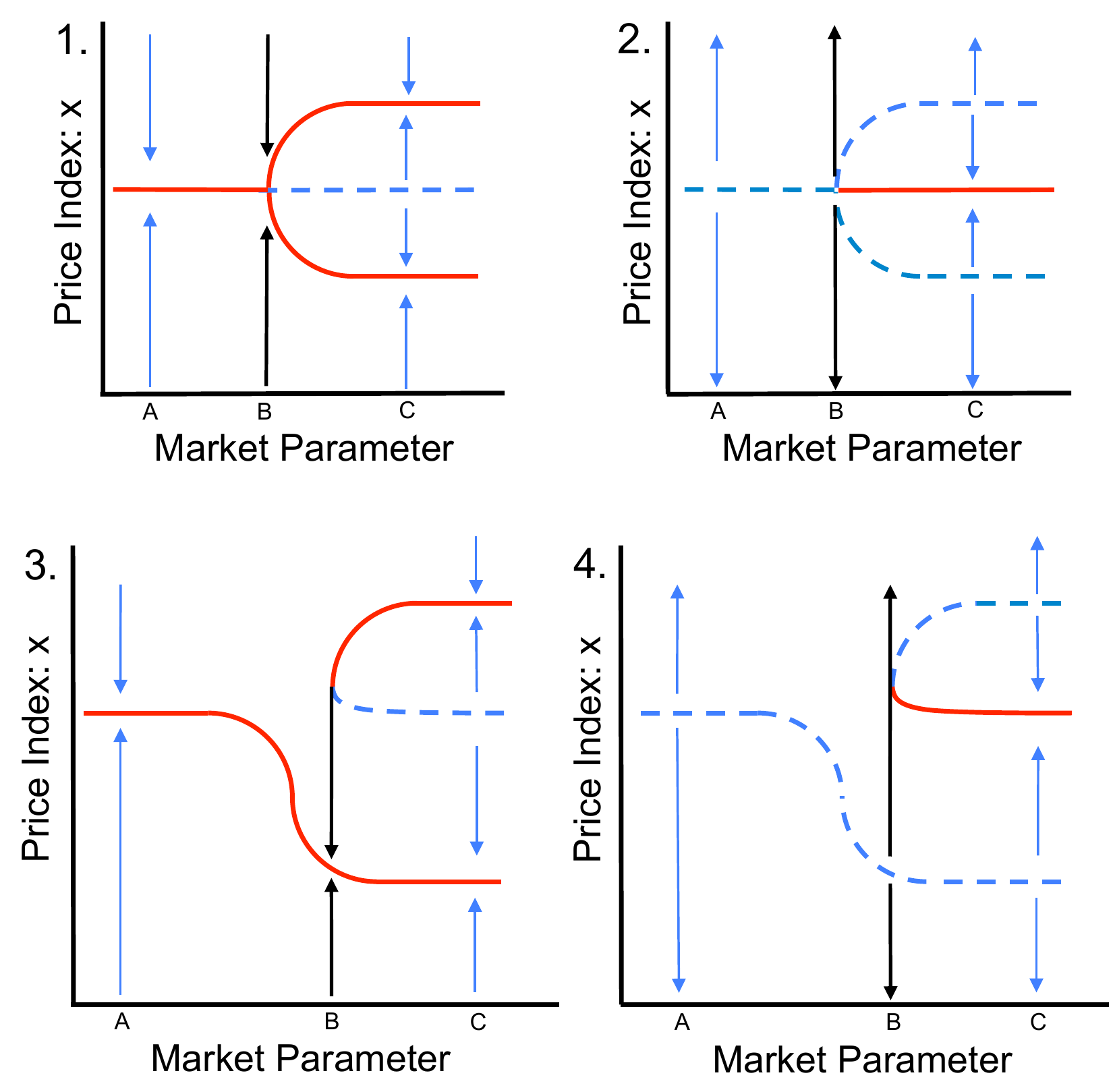}
\caption{\label{Multiple_Plots} {\bf Macroscopic properties of housing markets.} Four stylised representations of a housing market index in which the equilibrium price bifurcates at point $B$. Solid lines are stable equilibrium points of the market, dashed lines are unstable equilibrium points, and arrows indicate the direction the market would move if it were away from equilibrium. Bifurcations such as these have been derived from housing market data in~\cite{diks2016can}.}
\end{figure}

\section{Bifurcations in the Macro and Micro of Markets \label{section_background}}

Catastrophe theory arises naturally in optimisation problems~\cite[Chapt. 10]{arnol1992catastrophe} and Nash equilibria are an example in which optimisation leads to instabilities in the number of equilibria. In the first part of this section the stochastic generalisation of catastrophe theory is introduced. In the second part the relationship between instabilities in ordered preferences and instabilities in Nash equilibria are introduced. 

\subsection{Stochastic catastrophe theory and critical markets}

In the preface to {\it Catastrophe Theory}~\cite{arnol1992catastrophe} Arnol'd presents the earlier work in singularity theory that led to Thom developing catastrophe theory~\cite{thom1969topological}. The stability of singularities in an economic context correspond to the persistence of critical markets subject to minor perturbations in their description. Examples of catastrophe theory can be found in housing markets~\cite{wilson2011catastrophe} and the collapse of financial~\cite{zeeman1974unstable,barunik2009can,barunik2015realizing} and economic~\cite{diks2016can,harre2014strategic} markets. The approach begins with a potential function: $W(\textbf{x}|\textbf{u}): \mathbb{R}^{m\times n} \rightarrow \mathbb{R}^m$ in which $\textbf{x}$ is a vector of $m\leq 2$ state variables and $\textbf{u}$ is a vector of $n\leq 4$ control parameters~\cite{stewart1977catastrophe}. In a one dimensional system with state variable $x_t$ and control parameters $\mathbf{u} = [u_1,u_2]$ catastrophe theory examines the change in the number of stationary solutions of:
\begin{eqnarray}
dx_t & = & -\frac{\partial W(x_t|\mathbf{u})}{\partial x_t} dt \label{det_case}
\end{eqnarray} 
given by: $\frac{\partial W(x_t|\mathbf{u})}{\partial x_t} = 0$. For example the potential function: 
\begin{eqnarray}
W(x_t| \mathbf{u}) & = & \frac{1}{4}x_t^4 - \frac{1}{2}u_1x_t^2 - u_2x_t. \label{quart_util}
\end{eqnarray} 
has been used to describe asset markets~\cite{barunik2009can,diks2016can} in which $x_t$ is a market index. The stationary states of the system are the specific $x$ that solve the equation: 
\begin{eqnarray} 
\left.-\frac{\partial W(x_t | \mathbf{u})}{\partial x_t}\right|_{x_t = x} & = & -x^3 + u_1 x + u_2 \;\; =\;\;  0. \label{TES}
\end{eqnarray} 

\begin{figure}[!ht]
\center
\includegraphics[width=.95\columnwidth]{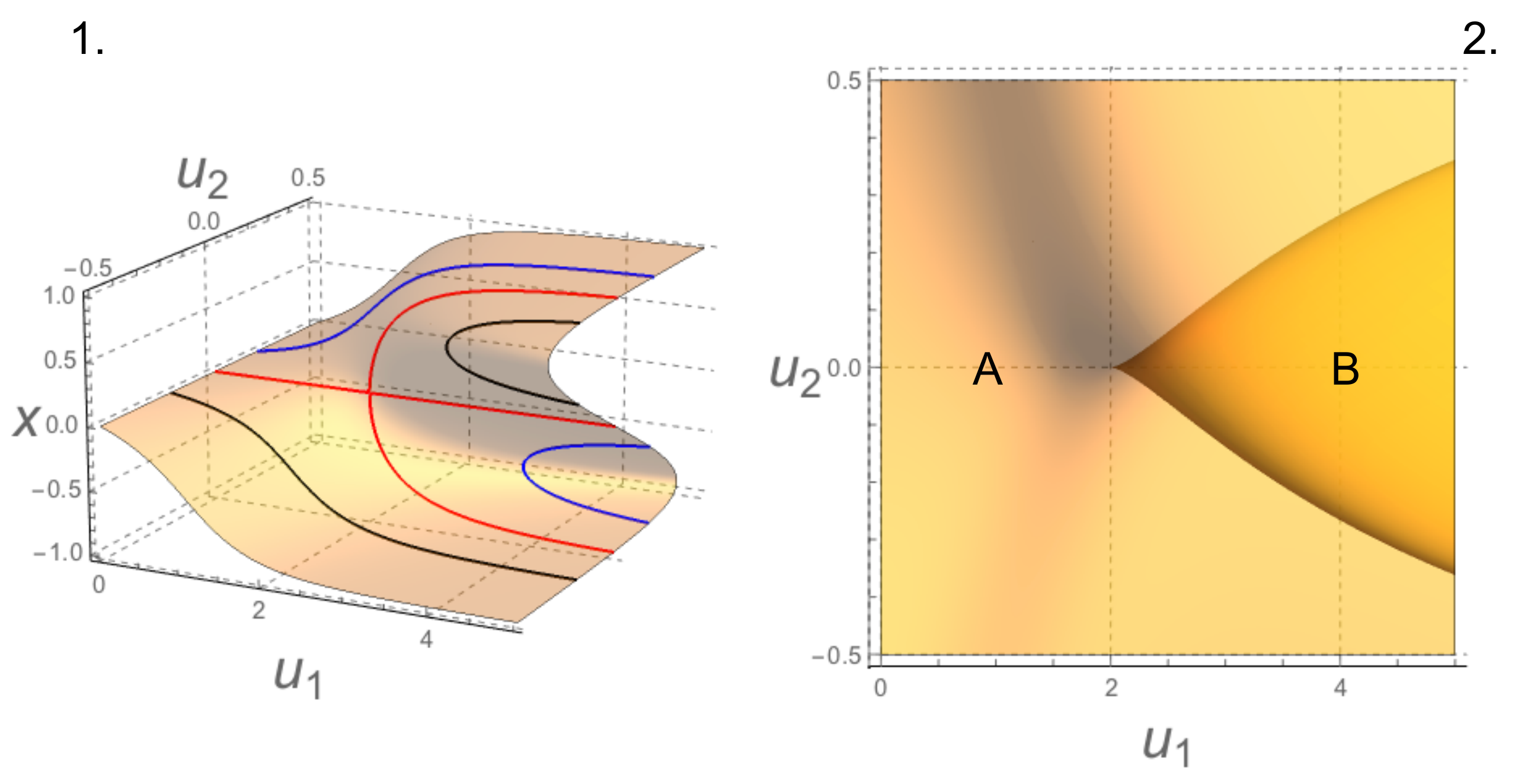}
\caption{\label{cusp_cat} Plot 1.: The {\it Cusp catastrophe} with three contours shown for fixed $u_2$ and varying $u_1$. A Pitchfork bifurcation is shown in red and two fold bifurcations are shown in black and blue. Plot 2.: Two distinct regions $A$ and $B$ can be discerned in the projection of the equilibrium surface onto the control plane $u_1 \times u_2$ and they are separated by the critical sets of $u_1 \times u_2$, region $A$ has one equilibrium point and region $B$ has three equilibrium points.}
\end{figure} 

Figure~\ref{cusp_cat} shows the time independent stationary points $x_t \equiv x$ of Equation~\ref{TES} as an equilibrium surface that is conditionally dependent on the control parameters $u_1$ and $u_2$. The covector $\mathbf{u}$ plays the same role as $m$ in the non-linear regression described in the Introduction, i.e. it is the parametric descriptor of a market. The projection of the bifurcation points onto the control plane $u_1 \times u_2$ is called the critical set and it separates region $A$ with one equilibrium from region $B$ with three equilibria. \\

A drawback of the original catastrophe theory is that stochasticity is not represented in the theory. Cobb~\cite{cobb1978stochastic} extended the theory to stochastic differential equations and this was improved on by Wagenmakers {\it et al.}~\cite{wagenmakers2005transformation}. The approach is to add noise to the evolutionary dynamics:
\begin{eqnarray}
dx_t & = & -\frac{\partial V(x_t|\mathbf{u})}{\partial x_t} dt  + \sigma(x_t) dW_t \label{SDE_cusp}
\end{eqnarray} 
in which $-\frac{\partial V(x_t|\mathbf{u})}{\partial x_t}$ ($\equiv \mu(x_t)$ for brevity) describes the deterministic evolution of the system, $dW_t$ is a Weiner diffusion process, and $\sigma(x_t)$ is the state dependent strength of the diffusion process. The stationary probability distribution of Equation~\ref{SDE_cusp} is~\cite{wagenmakers2005transformation}: 
\begin{eqnarray}
p(x|\mathbf{u}, \sigma(x)) & = & \mathcal{Z}^{-1}\exp\Big(2\int_{a}^{x_t} \frac{\mu(z) - 0.5(d_z\sigma(z)^2)}{\sigma(z)^2}dz\Big) \label{SDE_prob1} \\
p(x|\mathbf{u}, \xi) & = &  \mathcal{Z}^{-1}\exp\Big(\xi \int_{a}^{x_t} \mu(z)dz\Big)  \label{SDE_prob2} \\
& = & \mathcal{Z}^{-1}\exp\big(\xi V(x\,|\,\mathbf{u})\big)
\end{eqnarray} 
In Equation~\ref{SDE_prob1}: $d_z\sigma(z)^2 = \frac{d}{dz}\sigma(z)^2$, Equation~\ref{SDE_prob2} is a simplification in which $2\sigma(z)^{-2} = \xi \equiv $ constant in $z$, and $\mathcal{Z}^{-1}$ normalises the probability distributions. For the case in which the noise is independent of the state, $\sigma(x_t) \equiv \sigma$, the potential function of the deterministic case of Equation~\ref{det_case} is proportional to the potential function of the stochastic case of Equation~\ref{SDE_cusp}, i.e. $W(x_t | \mathbf{u}) = \xi V(x\,|\,\mathbf{u})$. With these assumptions the stationary points of the deterministic form coincide with the stochastic form. 

\subsection{Bifurcations in game theory}

In this section bifurcations in Nash equilibria are shown to be a result of the instability of agent preferences as game parameters vary. In the first part the binary preferences of an agent in the context of the decisions of other agents are introduced. In the second part instabilities in these preferences are shown to play a central role in the bifurcations of Nash equilibria. \\

Game theory has two interdependent components, the elements of the game agents play (number of choices, number of agents, utility functions) and the decision making (optimisation) process of the agents~\cite[Chapter 1]{camerer2003behavioral}, i.e. games provide a taxonomy of the strategic interactions between agents while game theory is a description of the basis on which agents optimise choices. Games also quantify the ordinal preferences of an agent (e.g. Gilboa~\cite[Chapter 7]{gilboa2009theory}) where preferences are a partially ordered binary relation $Q(\cdot, \cdot)$ over a choice set $X$: 
\begin{eqnarray}
Q(x_i,x_j) & = & x_i \succeq x_j, \,\, \,\, (x_i,x_j) \in X\times X,
\end{eqnarray} 
read as $x_i$ is weakly preferred to $x_j$. If $(x_i \succeq x_j)  \land (x_j \succeq x_i)$  an agent is indifferent between choices: $x_i \sim x_j$ and if $(x_i \succeq x_j ) \,\, \neg \,\, (x_i \sim x_j)$ an agent strongly prefers $x_i$ to $x_j$: $x_i \succ x_j$. Utilities extend ordinal preferences to real valued functions: if $x_j \in X$ a utility function is $U: X \rightarrow \mathbb{R}$ and we assume~\cite{gilboa2009theory} utilities represent ordinal preferences: $Q(x_i, x_j) \iff U(x_i) \geq U(x_j)$ $\forall i,j$. Games extend the utilities of an agent $i$ to utilities in the context of other agents' choices: given $x_j^i \in X^i$ and denoting the choices of all other agents as $x^{-i} \in X^{-i}$ then agent $i$'s game utility is $U_i: X^i\times X^{-i} \rightarrow \mathbb{R}$. We assume that $U_i$ represents $i$'s ordinal preferences $Q_i(x^i_j, x^i_k)$ in the context of the decisions $x^{-i}$ of other agents if $U_i(x^i_j,x^{-i}) \geq U_i(x^i_k,x^{-i})$  $\forall j,k$. The Nash equilibrium (NE) of an $n$ player game is an $n-$tuple of choices $(x_*^1, x_*^1, \ldots, x_*^n)$ such that:
\begin{eqnarray}
U_i(x_*^i, x_*^{-i}) & \geq & U_i(x_j^i, x_*^{-i}), \,\, i\, =\, 1, \ldots, n  \,\,\,\, \forall \,\,\,\, x_j^i \in X^i. \label{NE_basic}
\end{eqnarray}

In $2\times 2$ games (two players, two choices) there are two possible pure strategy outcomes\footnote{Mixed strategy NE are not important for what follows.}, either one or two NE (for the remainder of this section NE refers to pure strategy NE). The number of NE are dependent on each agent's preference relations $Q_i(\cdot, \cdot)$: if an agent's preference order changes then the number and location of the NE might also change. Consequently a parameterised utility function will have regions of the parameter space in which the preference relations of the agents change and that may result in qualitative changes in the NE. \\

\begin{figure}[!ht]
\center
\includegraphics[width=\columnwidth]{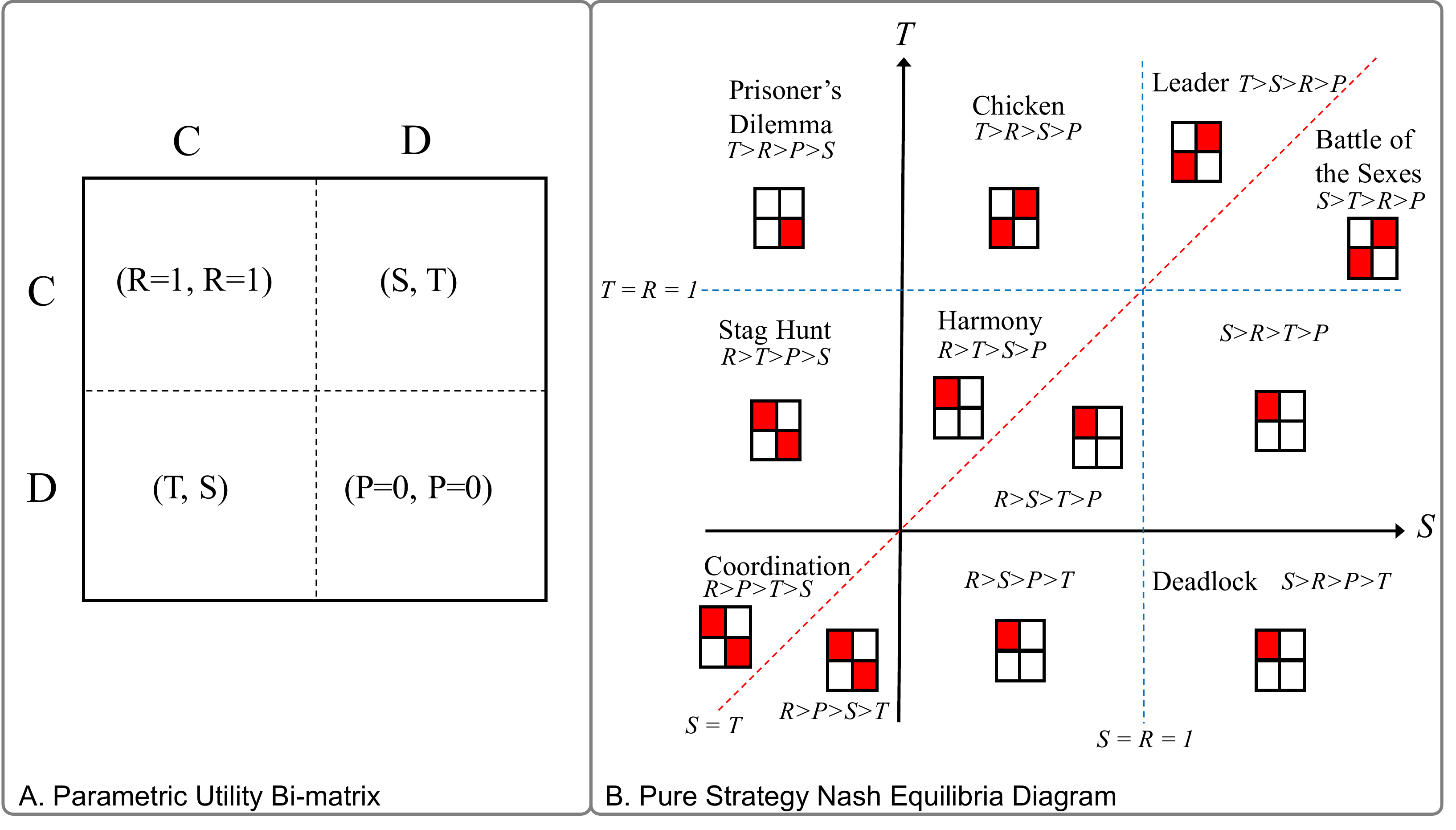}
\caption{\label{Nash_Eq_Bif_dia} {\bf Parametric games:} {\bf A.} The utility bi-matrix with two free parameters $T$ and $S$ for some symmetrical games. {\bf B.} The bifurcations of NE as $T$ and $S$ vary. The locations (strategy pairs) of the NE are shown as red boxes, e.g. the Harmony game has one NE:  ($CC$) and the Stag Hunt game has two NE: ($CC$), ($DD$).}
\end{figure}

An example of parameterised $2\times2$ games~\cite{shutters2013towards,poncela2016humans}\footnote{Rapoport and Guyer~\cite{rapoport1978taxonomy} find a total of 78 uniquely identifiable 2$\times$2 games.} is the utility bi-matrix shown in Figure~\ref{Nash_Eq_Bif_dia} A. in which $C$(ooperate) and $D$(efect) are the two choices available to the two agents. The utility for strategy ($C,C$) is $R=1$, for ($D,D$) is $P=0$ and these are held fixed while $T$ and $S$ are allowed to vary. In Figure~\ref{Nash_Eq_Bif_dia} B. the games are labelled with their common names (if they have one) and the ordinal relationships between $T$ and $S$ are shown for each region of the $T \times S$ parameter space. Note that not all changes in ordinal relationships results in a change in the number or location of NE. Looking at the relationship between the Stag Hunt, Harmony, Prisoner's Dilemma, and Chicken games a rich variety of changes in the NE occur as the preference ordering of $T$ and $S$ change. For example in going from the Stag Hunt game in which $P>S$ to the Harmony game in which $S>P$ a NE at $(D,D)$ disappears whereas going from the Harmony game to the Chicken game in which $R>T \rightarrow T>R$ the NE at $(C,C)$ disappears and two NE appear at $(D,C)$ and $(C,D)$. \\

Abrupt changes in the number of NE as utilities smoothly vary is characteristic of catastrophe theory in which optimisation leads to instability in the number of fixed points~\cite{arnol1992catastrophe}. In game theory this occurs when the agents' ordinal preferences become unstable as $Q_i(\cdot,\cdot)$ changes: $x_j \succ x_k \rightarrow x_k \succ x_j$ with an intermediate (critical) state in which $x_i \sim x_j$. The $2\times 2$ ordinal games with intermediate states have been characterised in~\cite{kilgour1988taxonomy}. In Section~\ref{QRE_intro} the Quantal Response Equilibrium will be introduced in which the Nash Equilibria (both pure and mixed) are a limiting case of the Quantal Response Equilibrium, and so the bifurcations of Figure~\ref{Nash_Eq_Bif_dia} are a limiting case of the bifurcations of the Quantal Response Equilibrium. Figure~\ref{game_cusp} is the generalisation to the Quantal Response Equilibrium of the Chicken game in Figure~\ref{Nash_Eq_Bif_dia} B.

\section{Three Models of Decision Theory \label{section_decision_theory}}

Catastrophe theory as presented above has a second drawback: it does not model the decisions of any economic agent, it is an aggregate (qualitative) model of market dynamics that cannot make quantitive predictions as pointed out by Thom~\cite{thom1977structural}. There are models though that are based on microeconomic decisions with bifurcations in their equilibrium states and that permit quantitive analysis. Here we present McFadden's~\cite{mcfadden1973conditional} original derivation and assumptions which are then extended to two other important classes of decision models. Some time is spent developing McFadden's model as the approach is similar to later models and can be used to shorten the work that follows. Brock and Durlauf's social decision model is based an decisions made in the context of social influences that modify an agent's utility. Similarly the Quantal Response Equilibrium is based on a stochastic extension of game theory that is shown to have a `social utility' component analogous to Brock and Durlauf's social utility function. It will also be shown that these two approaches give very similar probability distributions but the interpretations are somewhat different and highlight different aspects of microeconomic processes that can cause markets to collapse. For comparison with the stock-flow literature these models should be compared to the dynamic models of Gauldi {\it et al}~\cite{gualdi2015tipping} and Gallegati  {\it et al}~\cite{gallegati2011period} in which critical dynamics and market collapse readily appear.

\subsection{McFadden's approach to heterogeneity for discrete choices \label{McFaddens_Analysis}}

McFadden~\cite{mcfadden1973conditional} begins with an agent $i$ in a large market of $N$ heterogeneous agents in which each agent needs to choose an $x^i_j \in X^i$ where the value of each choice is given by a stochastic utility function:
\begin{eqnarray}
U(x^i_j \, | \, \mathbf{Z}^i,\varepsilon^i_j) & = & V(x^i_j \, | \, \mathbf{Z}^i) + \varepsilon^i_j \label{basic_DDE}\\
\textrm{Decision problem:}  \,\,\, x^i_j & = & \argmax{x^i_k \, \in \, X^i}(U(x^i_k \, | \, \mathbf{Z}^i,\varepsilon^i_k)). \label{DStoch_P}
\end{eqnarray}
$V(x^i_j \, | \, \mathbf{Z}^i)$ is a deterministic `public utility' that each agent knows. For example in a simplified housing market in which an agent chooses between two types of houses: $x^i_j \in \{ H_1, H_2 \}$ and $V(x^i_1 = H_1 \, | \, \mathbf{Z}^i)$ is conditional on the parameter vector $\mathbf{Z}^i$ of the property's characteristics that is particular to agent $i$ such as the number of bedrooms, land size, or build quality. The second component of Equation~\ref{basic_DDE} is the uncertainty in $U(x^i_j\, | \, \mathbf{Z}^i,\varepsilon^i_j)$ that has two common interpretations~\cite{brock2001discrete}. The first is that $\varepsilon^i_j$ is an innate psychological uncertainty such that each agent is only able to estimate $U(x^i_j\, | \, \mathbf{Z}^i,\varepsilon^i_j)$. The second is that $U(x^i_j\, | \, \mathbf{Z}^i, \varepsilon^i_j)$ is known precisely to agent $i$ but the population of other agents (or an economic modeller external to the market) only knows the public component $V(x^i_j\, | \, \mathbf{Z}^i)$ so that $\varepsilon^i_j$ is the uncertainty attributable to the limited knowledge an observer of agent $i$ has of their utility. In either case $\varepsilon^i_j$ is an idiosyncratic stochastic variable measured across the whole market so that the market is composed of agents that are heterogeneous in their utilities.\\

The decision problem of Equation~\ref{DStoch_P} is simplified by assuming all agents have identical utilities $V(x^i_j \,|\, \mathbf{Z}^i) \equiv V(x_j \,|\, \mathbf{Z})$ for all $x_j \, \in \, X$ (all $i$ superscripts are dropped for the remainder of this section). The complement set of a decision variable $x_{j}$ is $x_{-j} =  X \setminus x_j$. Following McFadden~\cite{mcfadden1973conditional} the probability that $x_j$ is a better choice than any other is:
\begin{eqnarray}
p(x_j \, | \, \mathbf{Z}, \varepsilon_j) & = & p(U(x_j\, | \, \mathbf{Z},\varepsilon_j) > U(x_k\, | \, \mathbf{Z},\varepsilon_j), \,\, \forall  \,\, x_k \in x_{-j}) \label{p_util} \\ 
& = & p(V(x_j\, | \, \mathbf{Z}) - V(x_k\, | \, \mathbf{Z}) > \varepsilon_k - \varepsilon_j, \,\, \forall  \,\, x_k \in x_{-j}). \label{p_util_noise} 
\end{eqnarray}
A useful assumption is that the cumulative distribution function (cdf) of the noise $\varepsilon = \varepsilon_j - \varepsilon_k$ is Gumbel distributed $\varepsilon \sim G(\mu, \gamma)$: 
\begin{eqnarray}
F(\varepsilon) & = & \exp(-e^{-(\varepsilon-\mu)/\gamma}), \label{cumm_DF}
\end{eqnarray}
where $\mu$ and $\gamma$ are non-negative parameters and the standard deviation is a function only of $\gamma$: $\sigma = \gamma \pi 6^{-1/2}$. Assuming the Independence of Irrelevant Alternatives (IIA) axiom~\cite{mcfadden1973conditional} $F(\varepsilon)$ is the correct cumulative distribution and not just a useful assumption. Dropping the $\mathbf{Z}$ notation for clarity, the probability distribution function (pdf) over choices is the logit distribution, which can be seen by rewriting Equation~\ref{p_util}:
\begin{eqnarray}
p(x_j \, | \, \varepsilon_j) & = & p(V(x_j) - V(x_k) + \varepsilon_j > \varepsilon_k, \,\, \forall  \,\, x_k \, \in \, x_{-j}) \\
& = & \int_{-\infty}^{+\infty} f(\varepsilon) \prod_{x_k \, \in \, x_{-j}} \exp\big(-e^{-(V(x_j) - V(x_k) + \varepsilon_k))/\gamma}\big) d\varepsilon, \label{pdf_int}
\end{eqnarray}
in which the location parameter of Equation~\ref{cumm_DF} has been shifted to $\mu =0$ and $$f(\varepsilon) = \gamma^{-1}\exp(-\varepsilon/\gamma-e^{-\varepsilon/\gamma})$$ is the pdf of $F(\varepsilon)$. Changing variables to $z = \exp(-\varepsilon/\gamma)$:
\begin{eqnarray}
p(x_j \, | \, \gamma) & = & \int_{0}^{+\infty} e^{-z} \prod_{x_k \, \in \, x_{-j}} \exp\big(-ze^{-(V(x_j) - V(x_k))/\gamma}\big) dz \\
& = & \int_{0}^{+\infty} e^{-z} \exp\bigg(-z\sum_{x_k \, \in \, x_{-j}}e^{-(V(x_j) - V(x_k))/\gamma}\bigg) dz \\
& = & \frac{1}{1+e^{-V(x_j)/\gamma}\sum_{x_k \, \in \, x_{-j}}e^{V(x_k)/\gamma}}\\
p(x_j \,|\, \mathbf{Z}, \xi) & = &  \frac{e^{\xi V(x_j \, | \, \mathbf{Z})}}{\sum_{x_k \, \in \, X }e^{\xi V(x_k \, | \, \mathbf{Z})}} \,\,\, \forall \,\,\, x_j \in X, \label{final_pdf}
\end{eqnarray}
where in the last step the conditional terms have been included again. When $\xi \rightarrow 0$ the estimate of $V(x_j)$ is dominated by the noise term $\varepsilon_j$ and $p(x_j) $ is uniform. When $\xi \rightarrow \infty$ there is no heterogeneity in the market  because $V(x_j)$ is identical for every agent and if there is a single objectively best choice $x_k$ then all agents will make the same choice with probability $p(x_k) = 1$. The decision problem of Equation~\ref{basic_DDE} then reduces to:
\begin{eqnarray}
x_k &=& \argmax{x_j \, \in \, X}(V(x_j \, | \, \mathbf{Z}))
\end{eqnarray}
which is deterministic and can also be seen by rewriting Equation~\ref{p_util}: 
\begin{eqnarray}
p(x_k \, | \,\mathbf{Z}) & = & p(V(x_k \, | \,\mathbf{Z})  > V(x_j \, | \, \mathbf{Z}) \,\, \forall \,\, x_j \in x_{-i} ) = 1.
\end{eqnarray}

In McFadden's model the choice of one agent has no influence on the outcome for any other agent. In this scenario the only way in which a critical market can occur is if $V(x_i\, | \,\mathbf{Z})$ in Equation~\ref{basic_DDE} is a high order polynomial, for example fourth order as in Equation~\ref{quart_util}, and the vector of coefficients $\mathbf{Z}$ is allowed to vary. \\ 

An illustrative example is given by McFadden in~\cite{mcfadden2001economic}. He describes $V(x_i\, |\, \mathbf{Z})$ for a single agent's decision problem as the {\it systematic utility} and implemented it as a linear function of measured attributes in which the elements of $\mathbf{Z}$ are the coefficients of a linear regression. The example he used is that of the econometric analysis of alternative freeway routes in which the regressors were observable attributes such as construction costs, route length, and parkland space and the coefficients of the regression reflected the `tastes' of decision-makers.

\subsection{Social Decision Theory \label{BDM_intro}}

This section develops social decision theory (SDT) by extending McFadden's Equation~\ref{DStoch_P} to include a social utility function. The approach follows Brock and Durlauf~\cite{brock2001discrete} in which social factors influence an agent's utility and consequently their decisions. \\

As before and following~\cite{durlauf2001interaction} an economic market is made up of $N$ agents and each agent has a value function that describes their pay-off for making a binary choice $x^i_j \in \{-1,1\}$. Agent $i$'s choice of $x^i_j$ maximises a utility function given a set of parameters that describe the characteristics of the agent's choices, these characteristics are represented by a vector of traits $\mathbf{Z}^{i}$ in the same way as described for Equation~\ref{basic_DDE}. Here we extend the earlier discussion by introducing a local influence from other agents. If $-i$ is the set of neighbouring agents that influence $i$'s utility for choice $x_j^i$, then a single neighbour of $i$ is $k \in -i$. Discrete choices are identified with a discrete numerical support: $x^i_j \in \{-1,1\}$ ($x^i_1 = -1$) so that expectations over decision variables have continuous support $\langle x \rangle^i =1-2p(x^i_1) \in [-1,1]$ and the vector of expected values: $\langle \bm{x} \rangle^{-i} \in [-1,1]^m$ is of length $m$. \\

An agent $i$ maximises a stochastic utility by making a binary choice $x^i_j \in \{1,-1\}$:
\begin{eqnarray}
U(x^i_j, \langle \bm{x} \rangle^{-i} \, | \, \mathbf{Z}^i, \varepsilon^i_j) & = & V(x^i_j \, | \, \mathbf{Z}^i) + S(x^i_j, \langle \bm{x} \rangle^{-i} \, | \, \mathbf{Z}^i ) + \varepsilon^i_j \label{dp_Utility} \\
\textrm{Decision problem:}\,\,\, x^i_j & = & \argmax{x^i_l \in \{-1,1\}} U(x^i_l, \langle \bm{x} \rangle^{-i} \, | \, \mathbf{Z}^i, \varepsilon^i_l).  \label{Optim_V}
\end{eqnarray}
This is an extension of Equations~\ref{basic_DDE} and~\ref{DStoch_P} in which $\langle \bm{x}\rangle^{-i}$ are the subjective expectations agent $i$ has of the strategies of neighbouring agents and $S(x^i_j, \langle \bm{x} \rangle^{-i}\, | \, \mathbf{Z}^i)$ is the social portion of the utility function representing interactions between $i$ and its neighbours. Brock and Durlauf~\cite{brock2001discrete} take the deterministic term $V(x^i_j \, | \, \mathbf{Z}^i)$ as linear in $x_j^i$ and choose the parameters of the following linear functions $h$ and $k$ so that $h$ is the difference between the deterministic parts of the public utility for the two choices:
\begin{eqnarray}
V(x^i_j \, | \, \mathbf{Z}^i) & = & k(\mathbf{Z}^i) + x^i_jh(\mathbf{Z}^i) \label{objective} \\
h(\mathbf{Z}^i) & = & \frac{1}{2}\big( V(1 \, | \, \mathbf{Z}^i) - V(-1\, | \, \mathbf{Z}^i) \big). \label{value_diff}
\end{eqnarray}
The social utility function is:
\begin{eqnarray}
S(x^i_j, \langle \bm{x}\rangle^{-i}  \,|\, \mathbf{Z}^i) & = & \sum_{k \in -i} J(\mathbf{Z}^i,\mathbf{Z}^k)x^i_j\langle x \rangle^k.
\end{eqnarray}
We simplify $J(\mathbf{Z}^i,\mathbf{Z}^k) = J^{i,k}$ and the total utility is:
\begin{eqnarray}
U(x^i_j, \langle \bm{x} \rangle^{-i} \, | \, \mathbf{Z}^i, \varepsilon^i_j) & = & k^i + x^i_jh^i + \sum_{k \in -i} J^{i,k}x_j^i \langle x\rangle^k + \varepsilon^i_j. \label{full_value}
\end{eqnarray}
Assuming the IIA and Gumbel distributed $\varepsilon^i_j$ the conditional probability over choices is:
\begin{eqnarray}
p(x^i_j = \pm 1\, | \, \langle \bm{x}\rangle^{-i} , \mathbf{Z}^i, \xi) & = &  \mathcal{Z}^{-1} \exp\Big(\xi x_j^i\big(h^i + \sum_{k \in -i} J^{i,k} \langle x \rangle^k \big) \Big), \label{Random_Probability_1}
\end{eqnarray}
where $\xi$ has the same interpretation as it did in Equation~\ref{final_pdf}. The $k^i$ and $h^i$ are the coefficients of an agent's public utility, and so they are known to all agents in the market ({\it c.f.} Equation~\ref{value_diff}). The average strength of interaction between $i$ and its neighbours $-i$ is $J^{i,-i}$ which replaces the summation over individual terms $J^{i,k}$ in Equation~\ref{full_value}. We interpret $J^{i,-i}$ as the strength of an endogenous `social field'. We also assume that the vector of strategies $\langle \bm{x}\rangle^{-i}$ can be approximated by $i$'s (subjective) average expected strategy of its neighbours: $\langle \bm{x}\rangle^{-i} = \langle x \rangle^{-i}$. Equation~\ref{Random_Probability_1} can be rewritten in the hyperbolic self-consistent form:
\begin{eqnarray}
\langle x \rangle^{i}_* & = &  \tanh\big( \xi (h + J^{i,-i} \langle x\rangle^{-i}_*) \big) \label{tanh_pdf}
\end{eqnarray}
In equilibrium an agents' strategies and their subjective expectations of the influence their neighbours have is the same: $\langle x\rangle^{i}_* \equiv \langle x\rangle^{-i}_*$, but maintaining this distinction is useful when discussing strategic complementarities and comparing models in Section~\ref{section_comp_analysis}. The public plus social plus private utility is:
\begin{eqnarray}
U(\langle x \rangle^{i}_*, \langle x \rangle^{-i}_*\, | \, \mathbf{Z}) & = & k + h\langle x \rangle^{i}_* + J^{i,-i}\langle x \rangle^{i}_* \langle x \rangle^{-i}_* + \varepsilon_j^i. \label{expected_value1}
\end{eqnarray}
where $\mathbf{Z} = [k,h,J^{i,-i}]$ is to be empirically estimated~\cite{brock2007identification}. Brock and Durlauf~\cite{durlauf2001interaction} observed that when $h = 0$ a pitchfork bifurcation occurs, e.g. Figure~\ref{Multiple_Plots} plot 1, as the uncertainty $\xi$ increases from 0 and when $h \neq 0$ a discontinuous bifurcation can occur, e.g. Figure~\ref{Multiple_Plots} plot 3. \\

There are also two possible contributions to the private utility $\varepsilon_j^i$ in Equation~\ref{expected_value1}, one from the public utility and the other from the social utility:
\begin{eqnarray}
\widetilde{V}(x^i_j \, | \, \mathbf{Z}^i) & = & V(x^i_j \, | \, \mathbf{Z}^i) + \overline{\varepsilon} \label{public_util_err} \\
\widetilde{S}(x^i_j, \langle \bm{x} \rangle^{-i} \, | \, \mathbf{Z}^i ) & = & S(x^i_j, \langle \bm{x} \rangle^{-i} \, | \, \mathbf{Z}^i)  + \underline{\varepsilon} \label{social_util_err} \\
\textrm{such that:} \,\,\varepsilon^i_j  & = & f(\overline{\varepsilon}, \underline{\varepsilon})
\end{eqnarray}
A simple observation with an interesting interpretation is that if $\varepsilon^i_j $ is attributable only to the heterogeneity in the social component: $ f(\overline{\varepsilon}, \underline{\varepsilon}) = \underline{\varepsilon} = \varepsilon_j^i$ then agent heterogeneity is due solely to the interaction term between agents while an individual agent comprehends their public utility perfectly well. For example in a housing market, $V(\cdot)$ represents the qualities of a house the buyer wants to buy, the `systematic utility' to which McFadden's regression analysis could be applied. However the influence that the house's neighbourhood has on either the buyer's or the seller's expectation of price, for example the number of houses recently sold in the neighbourhood and for how much, has an uncertain influence on the price the buyer expects to achieve: the uncertainty in the buyer's total utility is due solely to the uncertainty in the utility of the buyer-seller interaction.

\subsection{The Quantal Response Equilibrium \label{QRE_intro}}

The Quantal Response Equilibrium (QRE) is an alternative to social decision theory when payoffs are interdependent. Brock and Ioannides~\cite{durlauf2010social} have suggested that there is a close parallel between the Quantal Response Equilibrium and Social Decision Theory, here this connection is made explicit. \\

In $2\times 2$ games there are two agents, $\mathcal{P}^1$ and $\mathcal{P}^2$, playing a game $\mathcal{G}$ against each other, each with a binary choice given by two payoff matrices $\mathcal{G} = \{\mathcal{G}^1,\mathcal{G}^2\}$:
\begin{eqnarray*} \mathcal{P}^1: \;\;
 \begin{array}{c|rr}
 \mathcal{G}^1 & x^2_1 & x^2_2 \\ 
  \hline 
  x^1_1 & a^1 &  c^1 \\
  x^1_2 & b^1 & d^1  \\ 
 \end{array}
 &\;\;\;\;\;\;& \mathcal{P}^2: \;\;
 \begin{array}{c|rr}
 \mathcal{G}^2  & x^2_1 & x^2_2 \\
  \hline
  x^1_1 & a^2 &  c^2 \\
  x^1_2 & b^2 & d^2  \\ 
 \end{array} \label{Arb_Util_Mat}
\end{eqnarray*}
in which $\mathcal{G}^i$ denotes the utility function of $\mathcal{P}^i$. To align notation with SDT $\langle x \rangle^i =1-2p(x^i_1) \in [-1,1]$ and the expected utilities can be written:
\begin{eqnarray}
V(\langle x \rangle^1, \langle x \rangle^2 \, | \, \mathcal{G}^1) & = & g^1_0 + \langle x \rangle^1 g^1_1 + \langle x \rangle^2 g^1_2 + \langle x \rangle^1 \langle x \rangle^2 g^1_{1,2} \label{QRE_U1_utils}\\
V(\langle x \rangle^1, \langle x \rangle^2 \, | \, \mathcal{G}^2) & = & g^2_0 + \langle x \rangle^1 g^2_1  + \langle x \rangle^2 g^2_2 + \langle x \rangle^1 \langle x \rangle^2 g^2_{1,2} \label{QRE_U2_utils}
\end{eqnarray}
where the parameters are derived from the utility functions:
\begin{eqnarray}
g^i_0 &=& (a^i+b^i+c^i+d^i)/4, \label{g0} \label{QRE_param1} \\
g^i_1 &=& ((b^i+d^i) - (a^i + c^i))/4, \label{g1} \label{QRE_param2} \\
g^i_2 &=& ((c^i + d^i) - (a^i+b^i))/4, \label{g2} \label{QRE_param3} \\
g^i_{1,2} &=& ((a^i+d^i) - (b^i+c^i))/4. \label{g12} \label{QRE_param4}
\end{eqnarray}
The utilities are grouped to emphasise particular relationships: $g_0^i$ is the utility for a uniform strategy $\langle x \rangle^1,\langle x \rangle^2 = 0$, $g^i_1$ is the difference of the sum of row terms, $g^i_2$ is the difference of the sum of column terms, and $g^i_{1,2}$ is the difference of the sum of diagonal terms. The $\langle x \rangle^i$ weight these different parameters. The differences $g^i_1$ and $g^i_2$ are generalisations of Equation~\ref{value_diff} to that of two agents playing a game and $g^i_{1,2}$ is the agent to agent equivalent of the social field $J^{i,-i}$ of Equation~\ref{tanh_pdf}. These relationships are discussed in Section~\ref{struct_sim}. The conditional expected utilities are the utilities to agent $i$ given the choice of $-i$:
\begin{eqnarray}
V^i(\langle x \rangle^{-i} \, | \, \mathcal{G}^i, x_j^i = \pm 1) &=& g^i_0 \pm  g^i_i + \langle x \rangle^{-i} g^i_{-i} \pm \langle x \rangle^{-i} g^i_{1,2}
\end{eqnarray}
Using the conditional expected utilities McKelvey and Palfrey~\cite{mckelvey1995quantal} extended the Nash equilibrium to agents who are imperfect decision-makers by introducing an error vector for each of $i$'s choices $\mathbf{\varepsilon}^i = [\varepsilon^i_1, \varepsilon^i_2]$ that are Gumbel distributed as in Equation~\ref{cumm_DF}:
\begin{eqnarray}
U^i( x^i_j, \langle x \rangle^{-i} \, | \, \mathcal{G}^i, \varepsilon^{i}_j) & = & g^i_0 + x^i_j g^i_i + \langle x \rangle^{-i} g^i_{-i} + x^i_j \langle x \rangle^{-i} g^i_{1,2} + \varepsilon^i_j \\
z & = & V^i(\langle x \rangle^{-i} \,| \, \mathcal{G}^i, x^i_2 ) - V( \langle x \rangle^{-i} \, | \, \mathcal{G}^i,x_1^i) \label{WWW} \\
p(z > \varepsilon^i_2-\varepsilon^i_1) & = & \frac{1}{1+\exp(-\xi^i z)}. \label{XXX}
\end{eqnarray}
Given the IIA axiom and the approach used in Section~\ref{McFaddens_Analysis} the QRE decision model of McKelvey and Palfrey~\cite{mckelvey1995quantal} is:
\begin{eqnarray}
p(x^i_j \, | \, \mathcal{G}^i, \langle x \rangle^{-i}, \xi^i) & = & \mathcal{Z}^{-1} \exp\big(\xi^i V^i(\langle x \rangle^{-i} \, | \, \mathcal{G}^i, x^i_j) \big) \label{QRE1}
\end{eqnarray}
Written in the equilibrium hyperbolic form of Equation~\ref{tanh_pdf}:
\begin{eqnarray}
\langle x \rangle^i_* & = &  \tanh\big( \xi^i (g^i_i + g_{1,2}^i \langle x \rangle^{-i}_*) \big) \label{tanh_QRE1} 
\end{eqnarray} 
In equilibrium SDT and the QRE include the subjective expectations of other agents $\langle x\rangle^{-i}$, but the SDT model in equilibrium has $\langle x\rangle^{i}_* \equiv \langle x\rangle^{-i}_*$ whereas this may not be true for the QRE, depending on each agent's individual parameters $\mathcal{G}^i$ and $\xi^i$. \\

To establish their result, McKelvey and Palfrey assumed the Gumbel distribution but not the IIA axiom. Instead they used Brouwer's Fixed Point Theorem and Sard's Theorem (Theorem 1 and the Appendix of~\cite{mckelvey1995quantal}) to establish the existence and discreteness of equilibria just as Debreu did in his original analysis of the existence and countability of equilibria for deterministic economies~\cite{debreu1970economies}. \\

Equations~\ref{WWW} and~\ref{XXX} illustrate how critical preferences of game theory ({\it c.f.} the discussion following Equation~\ref{NE_basic}) enter into probability distributions over utilities. In the case of zero uncertainty in utilities Equation~\ref{XXX} reduces to $p(V^i(\langle x \rangle^{-i} \,| \, \mathcal{G}^i, x^i_2 ) > V^i( \langle x \rangle^{-i} \, | \, \mathcal{G}^i,x_1^i)) = 1 \iff x_2^i \succ x_1^i$. Preference relations are critical when $x_2^i \sim x_1^i$ and bifurcations may occur in game theory illustrated in Figure~\ref{Nash_Eq_Bif_dia}.

\section{Comparative Analysis, Critical Markets, and Twin Crises \label{section_comp_analysis}}

\subsection{Structural similarities in mathematical descriptions \label{struct_sim}}

A comparison between the QRE, Equation~\ref{QRE1}, and the SDT, Equation~\ref{Random_Probability_1}, shows their structural similarities. An important distinction though is that in the QRE the interaction terms given by $\mathcal{G}^i$ can differ between agents whereas the SDT only considers interactions between agents of the same type. Writing the two utilities together to compare coefficients of like terms:
\begin{eqnarray}
\textrm{SDT:} \,\,\,\, U(\langle x \rangle^i,\langle x \rangle^{-i} \, | \, \mathbf{Z}^i, \varepsilon) & = & k^i + \langle x \rangle^i h^i + \langle x \rangle^i \langle x\rangle^{-i} J^{i,-i} + \varepsilon, \label{fin_util_1} \\
\textrm{QRE:} \,\,\,\, U( \langle x \rangle^i, \langle x \rangle^{-i} \, | \, \mathcal{G}^i, \varepsilon^{i}) & = & g^i_0 + \langle x \rangle^i g^i_i + \langle x \rangle^{-i} g^i_{-i} + \langle x \rangle^i \langle x \rangle^{-i} g^i_{1,2} + \varepsilon^i. \label{fin_util_2}
\end{eqnarray}
The coefficients $g^i_0$ and $g^i_{-i}$ appear in the QRE utilities but not the QRE itself (Equation~\ref{tanh_QRE1}) and the $k^i$ term appears in the SDT utility but not in the equilibrium solution of the SDT (Equation~\ref{tanh_pdf}). However they are important at the level of the utility function as follows. The spillover from agent $-i$ to agent $i$ for some deterministic utility function $V^i(x^{i}, x^{-i})$ with continuous decision variables $x^i$ and $x^{-i}$ is defined as~\cite{cooper1988coordinating}:
\begin{eqnarray}
\frac{\partial V^i(x^i, x^{-i})}{\partial  x^{-i}} = 
\begin{cases}
> 0 \textrm{ positive spillovers}  \\
< 0 \textrm{ negative spillovers}
\end{cases} \label{spillover}
\end{eqnarray}
Assuming for simplicity $J^{i,-i} = g_{1,2}^i = 0$ so that there are no interaction terms in agent utilities, then the SDT utility has no spillover from $-i$ to $i$, whereas the spillover for the QRE utility is:
\begin{eqnarray}
\frac{\partial V(\langle x \rangle^i, \langle x \rangle^{-i} \, | \, \mathcal{G}^i)}{\partial  \langle x \rangle^{-i}} & = & g_{-i}^i
\end{eqnarray}

A notable example of spillovers is in the Prisoner's Dilemma in which $g_{1,2}^i =0$ and negative spillovers dominate the payoff to both agents, i.e. $g_{-i}^i > g_i^i$. In this case not only are the QRE and the SDT models indifferent to spillovers at the level of decision-making, but so is the utility of the SDT (but not the QRE). This can be accounted for by extending the SDT utility:

\begin{eqnarray}
\textrm{SDT:} \,\,\,\, U(\langle x \rangle^i,\langle x \rangle^{-i} \, | \, \mathbf{Z}^i, \varepsilon_j) & = & k^i + \langle x \rangle^i h^i + \langle x\rangle^{-i} f^i +  \langle x \rangle^i \langle x\rangle^{-i} J^{i,-i} + \varepsilon_j, \label{fin_util_3}
\end{eqnarray}

Then the extended vector of coefficients for the SDT: $\mathbf{Z}^i = [k^i, h^i, f^i, J^{i,-i}]$ are formally equivalent to the interactions between two agents playing an underlying game $\mathcal{G}$ if the game utilities are symmetric, for example the games in Figure~\ref{Nash_Eq_Bif_dia}. This equivalence can be seen by writing the payoff parameters as a vector $\mathcal{G}^i = [a^i, b^i, c^i, d^i]$ ({\it c.f.} Equations~\ref{QRE_param1} to~\ref{QRE_param4}) so that the relationship between $\mathcal{G}^i$ and $\mathbf{Z}^i$ can be written as a system of linear equations $A\mathcal{G}^i = \mathbf{Z}^i$:
\begin{eqnarray} 
\frac{1}{4} \left[ \begin{array}{rrrr}
1 & 1 & 1 & 1 \\
-1 & 1 & -1 & 1 \\
-1 & -1 & 1 & 1 \\
1 & -1 & -1 & 1 \\ 
\end{array} \right]
\left[ \begin{array}{l}
a^i \\ b^i \\ c^i \\ d^i \\
\end{array} \right]
& = &
\left[ \begin{array}{l}
g^i_0 \\ g^i_1 \\ g^i_2 \\ g^i_{1,2} \\
\end{array} \right]
\,\, = \,\, 
\left[ \begin{array}{l}
k^i \\ h^i \\ f^i \\ J^{i,-i} \\
\end{array} \right] \label{linear_BDM_QRE}
\end{eqnarray}
The QRE utilities can be recovered from the SDT parameters by solving $\mathcal{G}^i = A^{-1}\mathbf{Z}^{i}$:
\begin{eqnarray} 
\left[ \begin{array}{rrrr}
1 & -1 & -1 & 1 \\
1 & 1 & -1 & -1 \\
1 & -1 & 1 & -1 \\
1 & 1 & 1 & 1 \\ 
\end{array} \right]
\left[ \begin{array}{l}
k^i \\ h^i \\ f^i \\ J^{i,-i} \\
\end{array} \right]
& = &
\left[ \begin{array}{l}
a^i \\ b^i \\ c^i \\ d^i \\
\end{array} \right]
\end{eqnarray}
The SDT parameters can be found empirically via econometric methods, see for example~\cite{brock2001discrete,durlauf2010social} for recent work. This relationship between the two formulations is important because observing game theory utilities directly in economic data might be difficult whereas econometric techniques are known for SDT. The critical term is $J^{i,-i}$ as this coefficient encodes the strategic complementarity in the SDT~\cite{brock2007identification} and is necessary for bifurcations in the QRE and SDT, introduced next.

\subsection{Critical markets and strategic complementarity}

Spillover effects cannot cause critical markets by themselves. For critical markets there needs to be an interaction between the agents' at the level of their decisions and not just their utilities. The appropriate analysis is that of strategic complementarity and strategic substitutability~\cite{cooper1988coordinating}, the definition will be given and then the implications for the SDT and QRE are discussed. Using the notation of Equation~\ref{spillover} the definition is:
\begin{eqnarray}
\frac{\partial^2 V^i(x^i, x^{-i})}{\partial  x^{-i}\partial  x^{i}} = 
\begin{cases}
> 0 \textrm{ strategic complementarity}    \\
< 0 \textrm{ strategic substitutability}   
\end{cases} \label{complementarity}
\end{eqnarray}
Applying the definition to Equations~\ref{fin_util_1} and~\ref{fin_util_2}:
\begin{eqnarray}
\frac{\partial^2 U(\langle x \rangle^{i}, \langle x \rangle^{-i}\,|\, \mathbf{Z}^i, \varepsilon )}{\partial \langle x \rangle^{i} \partial \langle x \rangle^{-i}} & = & J^{i,-i} \\
\frac{\partial^2  U(\langle x \rangle^i, \langle x \rangle^{-i} \, | \, \mathcal{G}^i, \varepsilon )}{ \partial \langle x \rangle^i \partial \langle x \rangle^{-i}} & = & g^i_{1,2} 
\end{eqnarray}
In $2 \times 2$ games it is sufficient for there to be three Nash equilibria which is satisfied if the strict inequalities $g^i_i \pm g_{1,2}^i < 0 <  g^i_i \mp g_{1,2}^i$ hold, {\it c.f.} the exponent in Equation~\ref{tanh_QRE1}. \\

\begin{figure}[!ht]
\center
\includegraphics[width=.95\columnwidth]{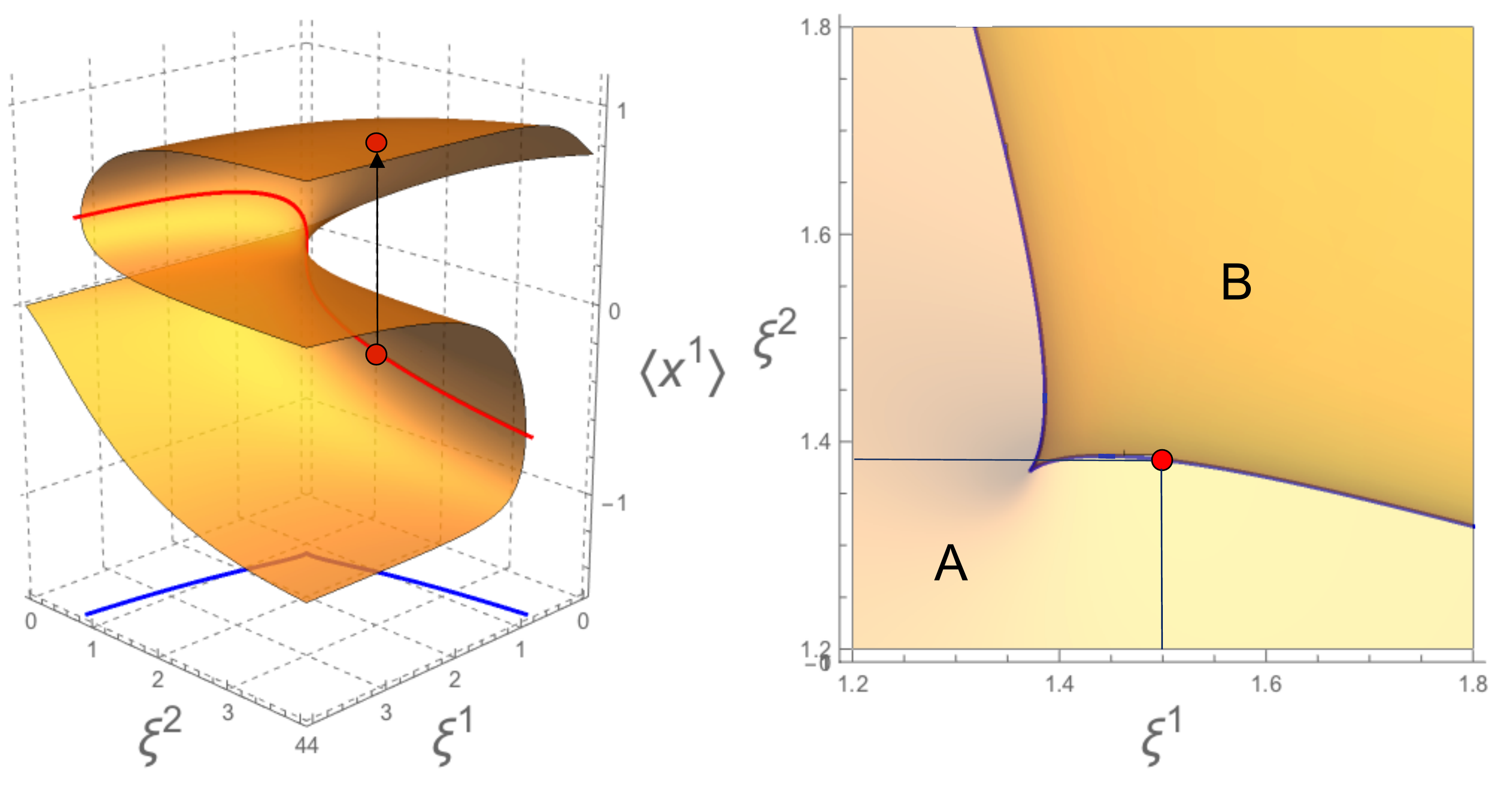}
\caption{\label{game_cusp} Left: The QRE for the Chicken game where the uncertainties $\xi^1$ and $\xi^2$ control the number of fixed points. The red curve is where the gradient of the fixed point surface diverges, defining the bifurcations at which the number of fixed points change. A critical market is shown with its transition from an unstable equilibrium to a stable equilibrium. Right: The critical set (blue curve) is the projection of the red bifurcation curve onto the control plane viewed from directly above, region A has one fixed point and region B has three fixed points. The surface, bifurcation curves, and critical set were derived using Mathematica.}
\end{figure}

To illustrate market criticality for the QRE consider a market in which agent-to-agent interactions are based on two types of agents, $\mathcal{P}^i$ and $\mathcal{P}^{-i}$, playing a game $\mathcal{G} = \{\mathcal{G}^i, \mathcal{G}^{-i} \}$ with fixed utilities. There are a total of $N$ (even) agents and there are $N/2$ agents of each type. The market operates by randomly assigning every agent of type $\mathcal{P}^i$ to pair with an agent of type $\mathcal{P}^{-i}$ and the payoff for their choices is given by $\mathcal{G}$. All agents of type $\mathcal{P}^i$ have an aggregate average strategy of $\langle x \rangle^{i}$ and likewise the aggregate average strategy of type $\mathcal{P}^{-i}$ agents is $\langle x \rangle^{-i}$. This model is based on the behavioural experiments using game theory across a population of players in~\cite{poncela2016humans}. The market uncertainties $\xi = [\xi_1, \xi_2]$ are allowed to vary and these uncertainties are described in more detail below. The example has symmetric utilities:
\begin{eqnarray*} \mathcal{P}^1: \;\;
 \begin{array}{l|cc}
 \mathcal{G}^1& x^2_1 & x^2_2 \\
  \hline 
  x^1_1 & 0 &  7 \\
  x^1_2 & 2 & 6  \\ 
 \end{array}
 &\;\;\;\;\;\;& \mathcal{P}^2: \;\;
 \begin{array}{l|cc}
 \mathcal{G}^2  & x^2_1 & x^2_2 \\
  \hline
  x^1_1 & 0 &  2 \\
  x^1_2 & 7 & 6  \\ 
 \end{array} \label{Chick_Util_Mat}
\end{eqnarray*}
There are three Nash equilibria, two pure: $(x^1_2,x^2_1)$, $(x^1_1,x^2_2)$ and one mixed: $(\langle x \rangle^1, \langle x \rangle^2) = (\frac{1}{3}, \frac{1}{3})$. $\mathcal{P}^i$'s expected utility and QRE are: 
\begin{eqnarray}
V( \langle x \rangle^i, \langle x \rangle^{-i} \, | \, \mathcal{G}^{i}) & = & \frac{1}{4}(15 + \langle x \rangle^i + 11 \langle x \rangle^{-i} - 3 \langle x \rangle^i \langle x \rangle^{-i} ), \\
\langle x \rangle^i_* & = & \tanh(\xi^i (1- 3 \langle x \rangle^{-i}_*)),
\end{eqnarray}
in which $g^i_{1,2} = -3$. Note that $1- 3 < 0 < 1 + 3$ so the QRE has either one or three equilibria except at critical parameter values: $\xi^* = [\xi^{1,*}, \xi^{2,*}]$. The QRE equilibrium surface parametrised by uncertainties $[\xi^1, \xi^2]$ is shown in Figure~\ref{game_cusp} for $\mathcal{P}^1$'s equilibrium $\langle x \rangle^1$. The red curve (left plot) is the bifurcation set on the equilibrium surface and the projection of the bifurcation set onto the control plane $\xi^1 \times \xi^2$ is the critical set $\xi^*$ of the market. 

\subsection{Strategic interactions and the `twin crises' effect}

The twin crises effect is a problem caused by the interactions between two markets. Kaminsky and Reinhart~\cite{kaminsky1999twin} studied the relationship between co-occurring crises in currency markets and banking finding that there was a causative relationship from a crisis in banking to a crisis in currency markets. Goldstein~\cite{goldstein2005strategic} has shown that strategic complementarities ({\it c.f.} Equation~\ref{complementarity}) play a crucial role in generating a feedback cycle between these markets during a crisis. Here we extend this notion to QRE based markets by showing that a bifurcation generated in one market induces a bifurcation in a second market, even if the second market's parameters $\mathbf{Z}^i = [a^i, b^i, c^i, d^i, \xi^i]$ remain fixed. \\

Previously interactions between agents were restricted to agents within a single market. In the following agents will represent two distinct markets where one market might produce a mixture of two goods that are then sold into another market of buyers. The notation developed earlier for the QRE of a single market composed of two types of agents passes through to the analysis of two markets described as two agents interacting. The average strategy of all producers $\mathcal{P}^1$ is\footnote{Angle brackets representing average strategies have been dropped to reduce clutter in the notation.} $x^1 \in [-1,1]$ and sellers $\mathcal{P}^2$ is $x^2 \in [-1,1]$ and we express the market QREs as functions of market uncertainties $\xi^i$, $\xi^{-i}$ and strategies $x^i$, $x^{-i}$: 
\begin{eqnarray}
 x^i  & = & f^i( x^{-i} , \xi^i), \label{Qi1} \\
& = & f^i(f^{-i}( x^{i} ,\xi^{-i}), \xi^i), \\
 x^{-i}  & = & f^{-i}( x^{i} , \xi^{-i}), \label{Qi2}\\
& = & f^{-i}(f^{i}( x^{-i} ,\xi^i), \xi^{-i}). \label{Qi22}
\end{eqnarray} 
To make this explicit in terms of the QRE, we can express each market's equilibrium state as:
\begin{eqnarray}
x^i_* & = & \tanh(\xi^i (g_i^i + g_{1,2}^ix_*^{-i})) \\
         & = & \tanh\Big(\xi^i \Big(g_i^i + g_{1,2}^i \tanh(\xi^{-i} (g^{-i}_{-i} + g_{1,2}^{-i}x^i_*) ) \Big)\Big) 
\end{eqnarray}

Note that each market is independent of the other market's strategy but they are coupled via their uncertainties $\xi^i$ and $\xi^{-i}$ and their parameter sets $\mathcal{G}$. To find the set of critical points where the QRE surface diverges we compute the Jacobian, see~\cite{wolpert2011strategic}. For example $\frac{\partial x^i}{\partial \xi^i}$ is:
\begin{eqnarray}
\frac{\partial f^i}{\partial x^{-i}}\frac{\partial f^{-i}}{\partial x^i}\frac{\partial x^i}{\partial \xi^i} + \frac{\partial f^i}{\partial \xi^i} - \frac{\partial x^i}{\partial \xi^i} & = & 0 \\
f_1^i f_1^{-i} \frac{\partial x^i}{\partial \xi^i} + f_2^i - \frac{\partial x^i}{\partial \xi^i} & = & 0  \label{dummy} \\
\frac{f_2^i}{f_1^i f_1^{-i} -1} & = &  \frac{\partial x^i}{\partial \xi^i} 
\end{eqnarray}
The simplification in notation at Equation~\ref{dummy} uses the subscript to denote differentiation with respect to either the first or the second argument of $f^i$ in equations~\ref{Qi1} and~\ref{Qi2}.  A similar set of computations for the remaining terms gives the Jacobian: 
\begin{eqnarray}
\mathbf{J}_{\xi}x & = & \frac{1}{f_1^i f_1^{-i} -1}
\begin{bmatrix}
    f_2^{i}       & f_1^{i}f_2^{-i} \\
    f_2^{i}f_1^{-i}       & f_2^{-i}
\end{bmatrix} \label{jacobian}
\end{eqnarray}
which diverges when $f_1^i f_1^{-i}=1$. The numerical solution of this equation has been mapped onto the equilibrium surface of Figure~\ref{game_cusp} (left plot) and then projected onto the control plane (right plot) to form the critical set of the market. \\

Note that the criteria for the equilibrium of market $i$ to be critical: $f_1^i f_1^{-i} =1$ is the same criteria for market $-i$ due to the symmetry between $i$ and $-i$, when the criteria is satisfied by one market it is satisfied by the other. Specifically, from Figure~\ref{game_cusp} if $\xi^1$ varies while $\xi^2$ remains fixed the equilibrium of both markets moves, markets can become critical markets and a market crisis will occur as $x^1$ discontinuously drops to a new and distant equilibrium point as illustrated by the arrow in Figure~\ref{game_cusp}. This induces a collapse in $x^2$, i.e. changes in the {\it uncertainty} in market $1$ causes a collapse in that market and this {\it induces a subsequent collapse} in market $2$ {\it despite market $2$ remaining unchanged} in its uncertainty $\xi^{2}$, causing twin crises to occur in the two markets. \\

To illustrate with a hypothesis, during the sub-prime mortgage crisis the financial markets may have changed their estimate of the value of the banks that held mortgages. So even if the banks knew the value of their own assets the change in the financial market's uncertainty in the value of the interaction between the two markets may have been sufficient to cause the twin crisis. This is intuitively sensible, if two markets are interconnected and one collapses this should have an impact on the equilibrium of the other market, which in turn has an impact on the first market, and so the market collapse feeds on itself. The general nature of this effect is reflected in the generality of Equations~\ref{Qi1} to~\ref{Qi22}. 

\section{Discussion \label{discussion}}


A shortcoming of the analysis presented here is that, apart from catastrophe theory, the models do not have a temporal aspect: all equations are time independent and it is implicitly assumed that each makes a decision at fixed point in time. This assumption is relaxed in Gallegati {\it et al}~\cite{gallegati2011period} based in part on the social interactions literature of Brock and Durlauf~\cite{durlauf2001interaction,brock2001discrete}. They use a stochastic model of a time dependent (log) price $p_t$ at time $t$ for a market with $N$ agents each having a binary decision variable $x_{i,t} \in \{-1,1\}$ and an excess demand function $x_{t-1}$ (notation is adapted to this article):
\begin{eqnarray}
x_{t-1} & = & N^{-1}\sum^{N}_{i=1}x_{i,t-1} \\
p_{t} & = & p_{t-1} + f(x_t) + z\sigma
\end{eqnarray}
where $f(x_t)$ is a deterministic function of the influence excess demand has on price evolution, $\sigma$ is an NID$(0,1)$ process, $z$ is the standard deviation and the equilibrium price $p^*$ is achieved for $f(x_t) = 0$, {\it c.f.} Equations~\ref{first_eg} and~\ref{SDE_cusp} for the time evolution of a market outcome or index for a known potential function. By introducing a static expectation that all agents have of the market's excess demand at time $t$: $x_t^e = x_{t-1}$ they write the time evolution of each agent's utility function as:
\begin{eqnarray}
U_{i,t} & = & (\overline{p}_t - p_{t-1})x_{i,t} + Jx_{i,t}x^e_{t} + \varepsilon_{i,t}
\end{eqnarray}
in which $\overline{p}_t$ is the expected (log) price at time $t$ and and the social interaction strength $J$ has the same interpretation as it does in the models discussed earlier. \\

The authors extend their model to include agents' wealth distribution dynamics, budget constraints, and positive transaction costs. These inclusions are sufficient to model the dynamics of market crashes that include a {\it period of financial distress} (PFD) in which a market peaks and might appear to have settled at a new equilibrium, or even be easing downwards, but this distressed period in fact precedes a market crisis. This period might last days or years. The perspective is microeconomic and described in terms of agent interactions~\cite[page 31]{kindelberger2005manias}: It is a period at which a large proportion of the agents are almost at the point of needing to liquidate assets as their ability to service debt contracts. Such PFDs are observed~\cite{gallegati2011period} in 36 of the 46 historical crises identified in Kindelberger's book {\it Manias, Panics and Crashes}~\cite{kindelberger2005manias} and it appears to be a key indicator of crises in asset markets. Some recent market dynamics, including the period immediately preceding the US sub-prime mortgage crisis and some sections of the current Australian housing market (at the time of writing), appear to have this PFD aspect to them. \\

The potential functions of catastrophe theory play a key role for whole market analyses in which there is a gradient dynamic~\cite{rosser2007rise}. In this case Equations~\ref{det_case} and~\ref{SDE_cusp} are integrable and the solution has the form of Equation~\ref{SDE_prob2}. In game theory this is true for two player, two choice games in which the mean valued strategies $\langle x\rangle^i $ reduce the number of strategic variables from two to one. This implies that the Brock and Durlauf binary model is integrable and so a potential function can be found, see Sandholm~\cite[Example 4.4]{sandholm2009large}. The requirement for a gradient system to be (locally) integrable is equivalent to symmetry of the Slutsky matrix of a market~\cite{afriat1977slutsky} and the relationship between the Slutsky matrices, potential functions, and catastrophe theory is covered in Balasko~\cite{balasko1978economic}. The general class of problems associated with potential functions and economic dynamics are covered in~\cite{varian1981dynamical} and Wagenmakers' analysis~\cite{wagenmakers2005transformation} allows econometric techniques to be applied to the detection of catastrophes. \\

A final note on the relationship between the different perspectives in this review. Gintis' call for game theory to unify the social sciences~\cite{gintis2009bounds} is consistent with econometric techniques used to detect social interactions between agents~\cite{durlauf2001interaction}. This in turn addresses Lucas' critique~\cite{brunner1983econometric} as the parameters that drive agent-to-agent interactions in game theory may be recovered from macroscopic measurements of collective social behaviour. This goes some way to understanding how the different approaches that have been developed in different branches of economic theory can be unified in a consistent way.

\section*{Acknowledgements}
This work was supported by Australian Research Council grant DP170102927.

\bibliographystyle{plain}

\end{document}